# Multimode vibrational strong coupling in Direct Laser written Mid-IR plasmonic MIM nano-patch antennas

Nicholas V. Proscia[1], Michael A. Meeker[1], Nicholas Sharac[1,¥], Frank K. Perkins[2], Chase T. Ellis[2], Paul D. Cunningham[2], Joseph G. Tischler,[2,†]

[1]*NRC Postdoc residing at US Naval Research laboratory, Washington, DC 20375, USA.*

[2]*U.S. Naval Research Laboratory, Washington, DC 20375, USA.*

[¥]*Current affiliation* Rigetti Computing, Freemont, CA 94538

[†]*Current affiliation Homer L. Dodge Department of Physics and Astronomy, The University of Oklahoma, Norman, OK 73019, United States*

Metal-Insulator-Metal (MIM) plasmonic structures can confine electromagnetic waves to a deep subwavelength regime, enabling strong light-matter interactions with potential applications in nonlinear optics and on-chip photonic circuitry. In addition, strong coupling of mid-infrared (mid-IR) vibrational transitions to optical cavities provides a way to modify and control a material's chemical reactivity and may also allow for highly sensitive chemical detection technology. Here, we experimentally and theoretically investigate the mid-IR optical properties of 3D-printed, nanoscale, anisotropic, L-shaped MIM plasmonic cavities. We observe strong vibrational-plasmon coupling between the two dipolar modes of the L-cavity and the polymer dielectric. The resulting three polariton modes are well described by a multimode coupled oscillator model, which we employ to predict the polariton behavior as a function of cavity arm length. The 3D printing technique offers time and cost reduction advantages over typical electron beam lithography and represents a highly accessible and versatile means of printing arbitrary-shaped nanometer-sized mid-IR plasmonic cavities capable of producing strong light-matter interactions for a variety of photonic or photochemical applications.

# Introduction

Surface plasmon modes hosted by noble metals, their nanostructures, and metamaterials are now routinely used to influence light-matter interactions. The resulting confinement and enhancement of local electromagnetic fields lead to surface enhancement of Raman scattering,[1] infrared (IR) absorption,[2] and nonlinear optical phenomena[3] while an increase in the vacuum density of states leads to Purcell enhancement of fluorescence.[4] Metal-Insulator-Metal (MIM) plasmonic structures, including MIM waveguides, nanoparticle-on-mirror, nanogap cavities, and patch antennas have emerged as a versatile platform for extreme subwavelength confinement of electromagnetic radiation, offering reduced losses and orders of magnitude stronger fields over a large bandwidth ranging from the visible to the Mid-IR.[5–8] These MIM structures work through mutual excitation of coherent localized surface plasmon modes on two metal interfaces separated by a thin dielectric layer leading to deeply subwavelength plasmonic modes based on Maxwell's equations. These modes have a lower damping rate than single-interfaced surface plasmon polaritons (SPPs),[9,10] and when one or two of the interfaces are finite in size, an ultra-small volume cavity can be created without a cutoff wavelength.[11] As a result of the several orders of magnitude increase in field strength, MIM structures offer a host of interesting applications related to enhancing light-matter interactions, which include use in chemical sensing,[12–14] quantum light sources,[15,16] lasers,[17] high-speed LEDs,[18] and exploiting chiral interactions with matter.[19,20]

When the light-matter interactions are large enough where the exchange of energy between the photonic and material systems occurs at a faster rate than the energy dissipation, this interaction between the systems is said to be in the strong coupling regime.[21,22] In this regime, the light and matter-state properties hybridize to produce new quasiparticles called polaritons[23] with novel behavior not present in either constituent system. The use of strong coupling to engineer new physical systems has proved invaluable in many areas of fundamental research.[24,25] While much of the past work has focused on modifying electronic transitions,[23,26–29] many recent efforts have been focused on coupling vibrational modes to photonic and plasmonic structures[30–35] to achieve vibrational strong coupling (VSC). The goals of such research include realizing ultrasensitive bio and chemical sensors[36], modifying chemical properties of materials, producing supramolecular assembly of polymers[37] and controlling thermal emissivity of material systems.[33,38–40] Due to the low oscillator strength of vibrational modes, it has been difficult to achieve VSC using localized surface plasmon resonances. Open plasmonic systems that are confined only in one or two directions are often employed because they can incorporate more oscillators to enhance the coupling strength, which scales as the square root of the number of oscillators, N.[41,42] To our knowledge, there has been only one report to date of plasmon-VSC in a closed plasmonic cavity system.[43] There, a MIM nanogap patch cavity design was employed due to its ultra-high field enhancement while maintaining reasonable Q. Such MIM structures are typically fabricated via time-consuming 2D electron beam lithography techniques and place limitations on the complexity of the three-dimensional structures that can be realized.

In this study, we use direct laser writing of 3D nanostructures to create L-shaped MIM nanogap cavities with fundamental resonances in the Mid-IR. This fast and inexpensive technique enables rapid prototyping and evaluation of photonic systems of arbitrary shape and complexity.[44,45] Previously, anisotropic L- and V-shaped plasmonic nanoparticles and their inverse apertured plasmonic films have proven to be a versatile platform for the manipulation of the near and far-fields of incident light, which include enhancement of birefringence,[46] wideband polarization conversion,[47,48] optical phased arrays,[49] as well as circular dichroism.[50] Here we apply this class of structure, with two orthogonal oscillator arms, in a nanogap MIM geometry thus confining the composite resonances primarily within the insulator volume. These L-shaped cavities are studied as a function of arm lengths and polarization angles. We find that the fundamental dipolar modes of each arm become an admixture of responses from both arms forming hybridized bonding and antibonding modes. Through the subwavelength field enhancement, we observe strong coupling between these localized hybrid dipolar modes of the L-cavity to the carbonyl (-C=O) stretch vibrational transition (CSVT) in the acrylate-based polymer insulator.[30] We model the system as a multimode three oscillator system between the two dipolar modes of the individual arms and the CSVT of the polymer insulator. Due to the large interaction potential between the two cavities modes, we find that three oscillator system experiences an extremely large splitting of 1101 cm$^{-1}$ between the upper and lower polariton branches. The large Rabi splitting of 104 cm$^{-1}$ between the lower and middle polariton branches reaches the strong coupling regime for mid-IR vibrational transitions of materials. As such, we establish direct laser written MIM structures as a promising platform on which to explore the effects of VSC in chemical and plasmonic systems. In particular, this fabrication technique could produce complex cavity geometries with nanoscopic voids in the insulator region for fluid flow to enable microfluidic plasmonic cavity devices for photochemistry applications.

# Methods and Materials

## Fabrication

The L-shaped nanogap cavities were fabricated via directing lasing writing, using a Nanoscribe Photonic Professional GT laser lithography system, where two-photon polymerization enables 3D printing of nanometer-sized structures.[44] A flow chart of the fabrication steps can be seen in Figure 1a. First, a 150 nm Au film was coated onto a silicon substrate by electron beam evaporation at pressures below $10^{-8}$ Torr, to ensure a high optical quality Au film thicker than the penetration depth, therefore fully reflective in the wavelength range of interest. Subsequently the Nanoscribe-proprietary photoresist, IP-Dip, was drop casted on top of the 150 nm Au film. 200 μm by 200 μm arrays of the same size L-shaped nanostructure with a constant 1600 nm pitch were written using the piezo mode, which corrects for tilting of the substrate. Multiple arrays were printed in which the arm length of the L was varied to investigate its effect on the resonance energy and polarization. After writing and development of the exposed IP-Dip resist, a film of 70 nm Au was deposited by electron beam evaporation on top, creating a L-shaped MIM nanogap cavity. The shape and dimensions of the L-cavities are shown in Figure 1b. The arm length along each axis was independently varied from a nominal length of 600 nm to 1400 nm in increments of ~200 nm. The widths of the L structures were kept to a nominal width of $300 \pm 15$ nm. Their shapes and dimensions were confirmed using scanning electron microscopy (SEM) and atomic force microscopy (AFM), as shown in Figures 1c-d.

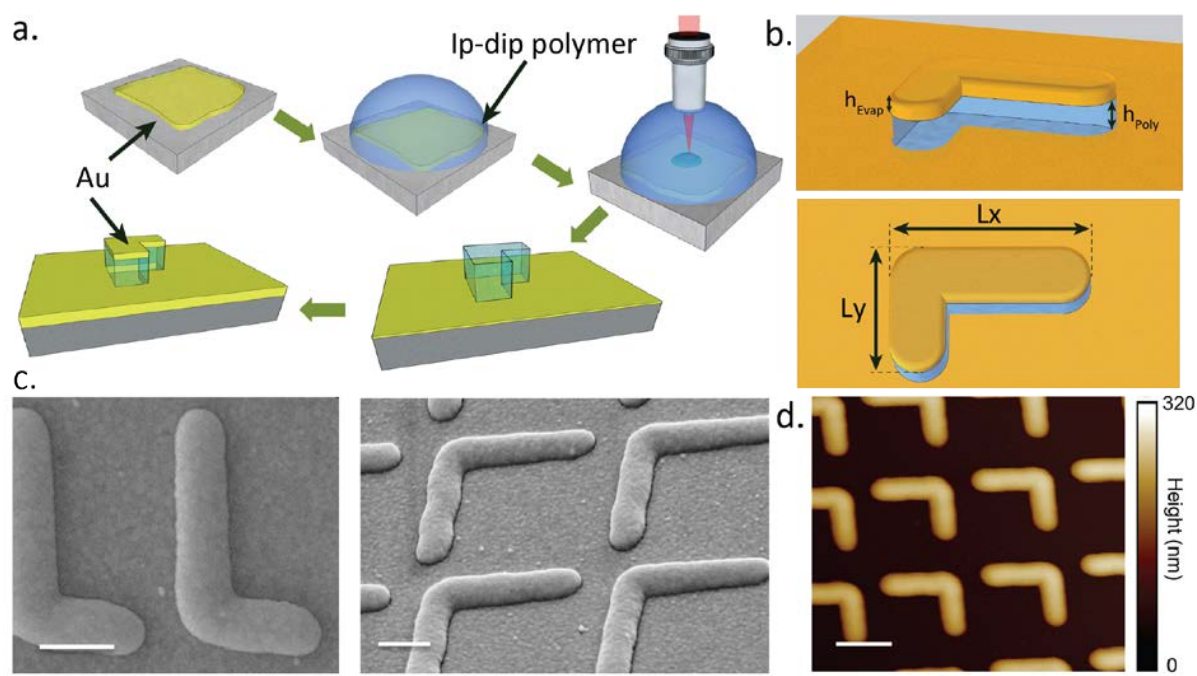

**Figure 1: Fabrication details & material characterization of the MIM L-cavities** (a) Flow diagram of the fabrication process involved in Direct Laser writing of the MIM L-cavities. (b) A graphic of the the final L- cavities used in EM simulations, Top: Angled side profile, Bottom: Angled overhead view. (c) SEM image of a fabricated L-cavity with arm lengths of 795 nm and 1324 nm (left). 45° image of similar L-cavities (right). Scale bars represent 400 nm. (d) AFM height sensor of a similar set L-cavities. Scale bar represents 1000 nm.

The final MIM structures are shown in the SEM and AFM images of Figure 1c-d and SI Figure S1, respectively. From the SEM images, we find that the periodicity (i.e. pitch) of the structures ranged from 1570 - 1610 nm where the nominal distance was 1600 nm. The nominal length of each arm was varied from 600 nm to 1400 nm in increments of 200 nm. The measured lengths were within 100 nm of the target length. Generally, the width of cavity arms ranged from 290 - 320 nm and decreased as the arm length increased. From the AFM image, the height of the developed polymer was 150 nm with a radius of curvature (ROC) of 50 nm along the top edges of the L-cavity. The ROC of the tips of L-cavities were approximately one half the width while the ROC of the outer intersection of the arms was found to be 150 nm.

### Optical Characterization

Polarized, mid-IR reflectance spectra were measured using an all-reflective, infrared microscope (Bruker Hyperion 1000) coupled to an FTIR spectrometer (Bruker Vertex 80v). The microscope is outfitted with a wire-grid polarizer and 36x magnification reflective objective that illuminates the sample with polarized light having an average incident angle of 22 degrees. The polarizer orientation is computer-controlled, and spectra were acquired with 5-degree increments of the polarization in order to obtain the polarization response of the measured samples.

### Numerical Simulations

3D, full-wave, electromagnetic (EM) simulations were performed using the finite element method (COMSOL Multiphysics – wave optics module). The simulation domain was constructed using periodic Floquet boundary conditions in conjunction with a periodic port for injection of plane wave excitation and a perfectly matched layer region below the substrate to absorb any light transmitted through the Au film. The angle of incidence (AOI) was fixed at 22º – consistent with measurements. The domain cross section was set to 1600 nm by 1600 nm corresponding to the nominal periodicity of the fabricated structures. The MIM L-cavities were approximated as rectangular planar films with the top corners of the evaporated Au and the Ip-polymer layer rounded by a ROC of 50 nm. The tips of the arms were rounded by 150 nm, in accordance with the experimental SEM and AFM images as shown in Figure 1c-d. The IP-Dip polymer film thickness was set to 150 nm. An image of a simulated L-cavity is shown in Figure 1b. The dielectric constants for IP-Dip, Silicon and Gold were taken from Refs [51], [52] and [53], respectively.

## Results

The optical response of a representative set of L-cavities is analyzed in Figure 2, where Figure 2c shows a series of polarization-resolved experimental reflectance spectra from an array of L-cavities with *Lx* = 797 and *Ly* = 1324 nm. Each reflectance dip in the spectra corresponds to the absorption of a distinct plasmonic resonance. The origin of each mode can be understood from the EM simulations of the structure. Figure 2 a & b show the out-of-plane electric field (Ez) profiles of the simulated modes at 50 nm above the initial Au substrate. The L-cavity modes consist of a series of out-of-plane dipolar and higher order resonances, labeled modes L, M, U, and H1-H4 in Figure 2 a-b. The resonant modes are created by in-phase LSPRs associated with the two interfaces created by the Au substrate and the 70 nm thick evaporated Au film on top of the Ip-dip polymer. The change in impedance at the edges of the structures confine the mode in the lateral dimensions [7]. For each of these modes, the electric field is primarily normal to the substrate (Ez) with significant in-plane (Ex and Ey) components only occurring from the fringe fields at the edges of the structure as shown in SI Figure S3.

The field mode plots in Figure 2a show the Ez-component of field distribution of the Lower (L), Middle (M) & Upper (U) polaritons. These are the dipolar modes of the cavity where modes U and L/M are orthogonal to each other, which is confirmed in the differential reflection polar plots of Figure 2d. This is consistent with a conductively coupled oscillator interpretation, where each arm of the cavity can be treated as an independent oscillator that couple together to produce orthogonal bonding and antibonding modes.[50] In this context, modes L and M take the character of the dipolar bonding mode, where the charge within each arm oscillates in-phase leading to destructive interference at the intersection of the two arms; while mode U has primarily an antibonding character, where the charges in both arms oscillate out-of-phase with each other creating constructive interference at the intersection of two arms. Modes L and M have a dipole direction of 110$^0$ relative to the x-axis as revealed in the field distribution (Figure 2a) and polarization plots (Figure 2d) whereas mode U is nearly orthogonal with a dipole moment oriented 22$^0$ relative to the x-axis. VSC between the dipolar plasmonic resonances of each individual arm and the CSVT of the IP-Dip polymer lead to splitting of what would have been the bonding and anti-bonding modes (see SI figure S7) into the lower, middle and upper polariton branches (modes L, M, and U). Modes L and M take the character of the bonding mode because, in the absence of VSC, the bonding mode would have formed close in energy to modes L and M. In the presence of VSC, the three oscillators hybridize and modes L & M inherit the bonding character while mode U is mostly of antibonding character.

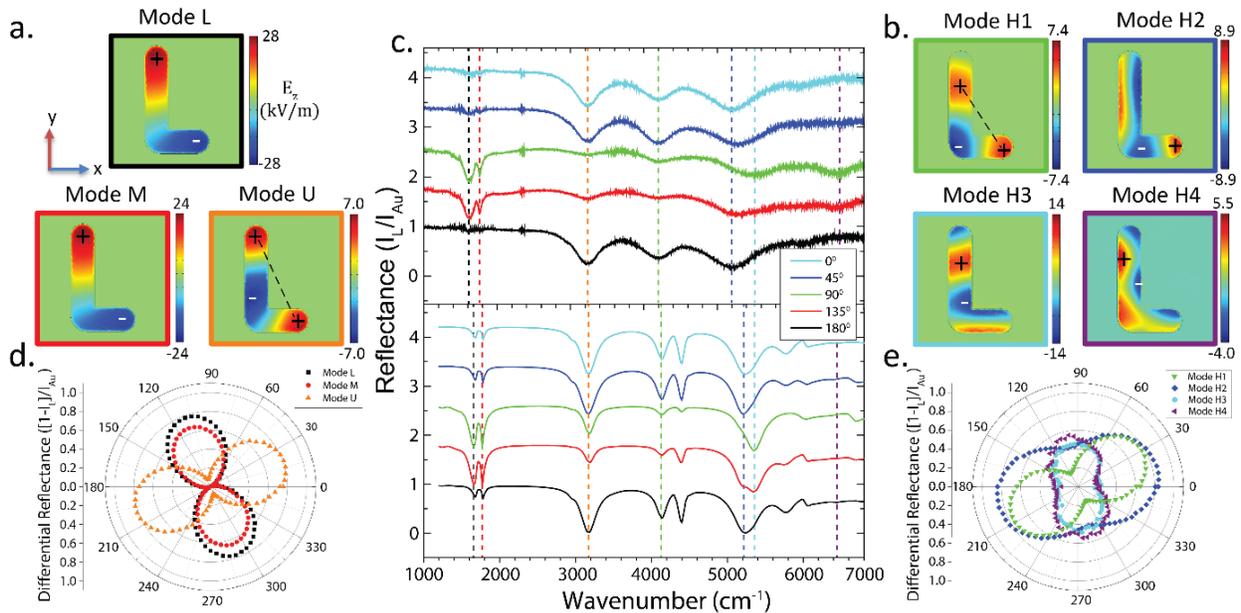

**Figure 2: Modal analysis of the L-cavity resonances** (a, b) The simulated out-of-plane electric field (Ez) profiles of the modes identified in the polarization-resolved reflectance spectra in (c). The modes are color-coded according to the border color of each plot. (c) The experimental (upper) and the simulated (lower) polarization-resolved reflectance spectra. The polarization angle of each measurement is indicated in the legend and the spectral locations of the modes are outlined by the color-coded dashed lines. The spectra in waterfall plots are separated by increments of 0.8. (d, e) Polar plots of the differential experimental reflectance of the modes shown in (a, b, c). $I_{Au}$ and $I_L$ are the reflected intensities from a planar Au film and L-cavity array with arm lengths of *Lx* = 797 nm and *Ly* = 1324 nm, respectively.

At larger wavenumbers, the resonances of the structure come from higher-order plasmon modes shown in Figure 2b for modes H1-H4. Mode H1 is another antibonding-like mode with an additional node at the tip of the Ly arm leading to a slight rotation in the dipole direction to 30$^0$ (Figure 2d). Mode H2 corresponds to a 2$^{nd}$ order resonance nearly along the x-axis, producing resonant field oscillations along the Ly arm width as well as the Lx arm length. Mode H3 is an asymmetric mode nearly orthogonal to mode H2, producing field oscillations along the width of the Lx arm as well as the Ly arm length. Mode

H4 is a superposition of higher-order resonances along both the x and y-directions. All localized modes of the L-cavities were independent of the incident angle of illumination, implying that scattering is the primary mechanism for coupling to the incident light.[43] There are several other higher order modes (at 4400, 5750 and 6050 cm$^{-1}$) that are not visible in the experimental reflectance data in Figure 2c (upper) but are predicted by the EM simulation in Figure 2c (lower). Of note is the mode at 4400 cm$^{-1}$ that is nearly degenerate with H1, and which may overlap with H1 in the experimental data due to linewidth-width broadening caused by defects in the fabricated cavities. The field profiles of these higher order modes are plotted in SI Figure S3 for completeness. These modes are not the focus of the rest of this report as they do not play a role in the light-matter coupling experienced by the CSVT and the cavity.

The multimode VSC in the system takes place between the two dipolar modes associated with each arm of the cavity (Figure S6c) and the CSVT of the IP-Dip polymer, resulting in three hybrid light-matter polariton modes. Figure 3a shows a waterfall plot of the experimental reflectance spectra of the three polariton modes for various Lx arm lengths. The Ly arm length was kept nearly constant with a mean length of 1300 nm while the Lx arm length was varied between 600 nm – 1400 nm in steps of ~200 nm. The measured arm lengths obtained from SEM images are displayed in the legend of Figure 3a. An avoided-crossing between modes L and M is clearly visible, with minimum energy difference between the branches for an *Lx* of 787 nm (figure 3b). Due to the symmetry of the L-shaped cavity, similar mode dispersions can be observed if *Ly* is varied instead of *Lx*, see SI Figure S4.

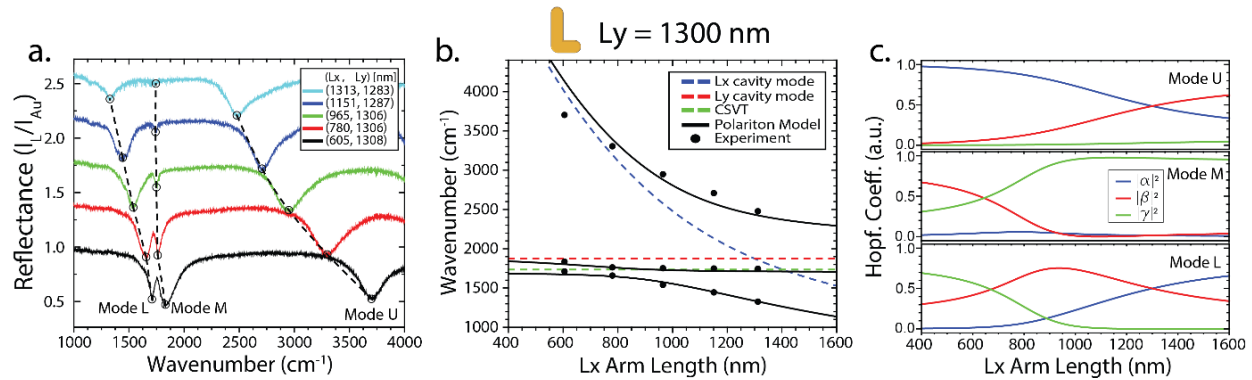

**Figure 3: Vibrational-Plasmon Strong coupling**. (a) Experimental reflectance of modes L, M & U in the frequency range where VSC to the IP-Dip vibrational mode occurs as the aspect ratio of the L is varied as a function of *Lx*. The dotted lines are guides to the eye indicating the polariton disperson. (b) Comparison between the measured location of the upper, middle and lower polaritons in (a) (black circles) and the coupled three oscillator model (black lines). The uncoupled *Lx* (blue) & *Ly* (red) modes predicted by Eq. 1 and the vibrational transition (green) are given by the dashed lines. (c) The hopfield coefficents or the eigenvectors for the eigenvalues of the three coupled oscillator model in Eq. (3) as a function of Lx arm Length.

In order to understand the formation of the polariton modes (L, M and U) of the system, we first need to comprehend the L-cavity modes in the absence of VSC. As described above, the L-cavity can be thought of as two coupled cavities.[54] Focusing on the lowest energy dipolar modes, where each arm of the cavity can be treated as an independent oscillator, we use a simple standing-wave condition to analytically model the arm modes as a function of length. Thus MIM gap-plasmon resonances are given by,

$$w \frac{2\pi}{\lambda} N_{eff} = m\pi + \phi, \quad (1)$$

where $w$ is the arm length of the cavity, $\lambda$ is the free space wavelength of the excitation, $m$ is the order of the mode, $\phi$ is the phase difference of the mode upon reflection at the termination boundaries, and $N_{eff}$ is the effective mode index given by $N_{eff} = k_{gap}/k_o$. For small polymer thicknesses, $t$, one can use the approximation $\tanh(x) \cong x$ where $x = \sqrt{k_{gap}^2 - \varepsilon_d^2 k_o} * \frac{t}{2}$. The wavenumber of the gap modes, $k_{gap}$, is then given by[11]

$$k_{gap} = k_0 \sqrt{\varepsilon_d + \frac{1}{2}\left(\frac{k_{gap}^0}{k_o}\right)^2 + \sqrt{\left(\frac{k_{gap}^0}{k_o}\right)^2 \left[\varepsilon_d - \varepsilon_m + \frac{1}{4}\left(\frac{k_{gap}^0}{k_o}\right)^2\right]}}, \quad (2)$$

where $k_o$ is the free-space wavenumber, $\varepsilon_d$ and $\varepsilon_m$ are the complex permittivity of the insulator and metal, respectively, and $k_{gap}^0 = -2\frac{\varepsilon_d}{t*\varepsilon_m}$. Following the method prescribed in Ref. [43] for modeling the uncoupled plasmonic modes, the Ip-dip Lorentzian and Gaussian transitions were removed from the $\varepsilon_d$ obtained from Eq. 1 in Ref [51]. The $\varepsilon_m$ for Au was obtained from Ref. [53]. The phase difference $\phi$ is dependent on the structural and material properties of the system and is related to extent of the plasmon field that extends past the edges of the cavity.[9] To determine $\phi$ for this nonplanar system, Eq. 1 is fitted to the EM simulated dispersion of a single arm resonance shown in SI Figure S6 with $\phi$ from Eq. 1 as a fitting parameter. The best fit to Eq. 1 is plotted in SI Figure S6c where $\phi = -7.33^0$. The resulting dispersion relation used to model the dipolar modes of the isolated arms Ly and Lx are shown by the red and blue dashed curves in Figure 3b. SI Figure S7 shows the expected hybrid bonded and antibonded cavity modes of the two coupled arms without the presence of Ip dip's CSVT. The hybrid modes were simulated via COMSOL and predicted with good agreement via a two oscillator strong coupling model of the two individual arm modes.

The VSC that gives rise to the polariton modes L, M and U can be described by a set of linear coupled equations between three oscillators: the dipolar cavity mode of the Lx arm, $\omega_x$, the dipolar cavity mode of the Ly arm, $\omega_x$, and the CSVT of the IP-Dip polymer, $\omega_{vib}$. The strongly coupled system is modeled in Figure 3b and 4 with the standard multimode coupled oscillator model given by an eigenvalue equation with the 3 x 3 matrix M,[35,55,56]

$$\begin{bmatrix} \omega_x[x] & V_1 & V_2 \\ V_1 & \omega_y & V_2 \\ V_2 & V_2 & \omega_{vib} \end{bmatrix} \begin{bmatrix} \alpha[x] \\ \beta[x] \\ \gamma[x] \end{bmatrix} = \lambda_{L,M,U} \begin{bmatrix} \alpha[x] \\ \beta[x] \\ \gamma[x] \end{bmatrix} \quad (3)$$

with eigenvalues, $\lambda_{L,M,U}$, given by $\det(\boldsymbol{M} - \lambda \boldsymbol{I}) = 0$,

$$(\omega_x[x] - \lambda_{L,M,U})[(\omega_y - \lambda_{L,M,U})(\omega_{vib} - \lambda_{L,M,U}) - V_2^2] - V_1[V_1(\omega_{vib} - \lambda_{L,M,U}) - V_2^2] + V_2[V_1 V_2 - V_2(\omega_y - \lambda_{L,M,U})] = 0. \quad (4)$$

here $\omega_x[x]$ is the dispersion of the plasmonic mode of the Lx arm given by Eq. 1 as a function of Lx arm length; $\omega_y$ the energy of the plasmonic mode for the Ly arm at average length of 1126 nm, $\omega_{vib}$ is the energy of the CSVT in wavenumbers. The coupling strengths are determined by $V_{1,2}$, which are the interaction potentials between modes $\omega_x[x]$ and $\omega_y$, and $\omega_x[x]$ or $\omega_y$ and $\omega_{vib}$ respectively. Given the uniform thickness of the cavity, $\omega_x$ and $\omega_y$ are assumed to have equal interaction potential to the $\omega_{vib}$, given by $V_2$. The eigenvectors $\alpha[x], \beta[x], \gamma[x]$ are the Hopfield coefficients of the polariton system associated with $\omega_x[x], \omega_y, \omega_{vib}$, respectively, and describe how much vibrational or plasmonic character is present in each polariton mode. By fitting the three eigenvalues, $\lambda_{L,M,U}$ to the measured resonances of each polariton branch, shown in Figure 3a and b, we find the interaction potentials $V_{1,2}$ to equal 541 cm$^{-1}$ and 82 cm$^{-1}$, respectively. From our three coupled oscillator model, we determine that Rabi splitting between polariton branches (modes) L-M, and U-L to be 104 and 1101 cm$^{-1}$, which occur for Lx arm lengths of 787 and 1290 nm, respectively. The Rabi splitting between polaritons M-U for *Ly* = 1300 occur at Lx arm lengths are outside the range of the model determined by the energy separation between $\omega_y$ and $\omega_{vib}$. Directly measuring the Rabi oscillations associated with these splitting values via coherent ultrafast spectroscopy is beyond the scope of the current study.

The Hopfield coefficients in Figure 3c show the weighted contribution of the original uncoupled modes to each polariton branch as a function of the Lx arm length. As the Lx arm increases in length, the U polariton goes from an almost purely $\omega_x$ mode to $\omega_y$ with a small contribution from $\omega_{vib}$. The M polariton goes from having a strong $\omega_y$ mode character to a mostly $\omega_{vib}$ where the $\omega_x$ mode remains a small contribution. While the L polariton evolves from primarily $\omega_{vib}$ character to a combination of the cavity modes and eventually approaching a purely $\omega_x$ nature at larger arm lengths (not shown).

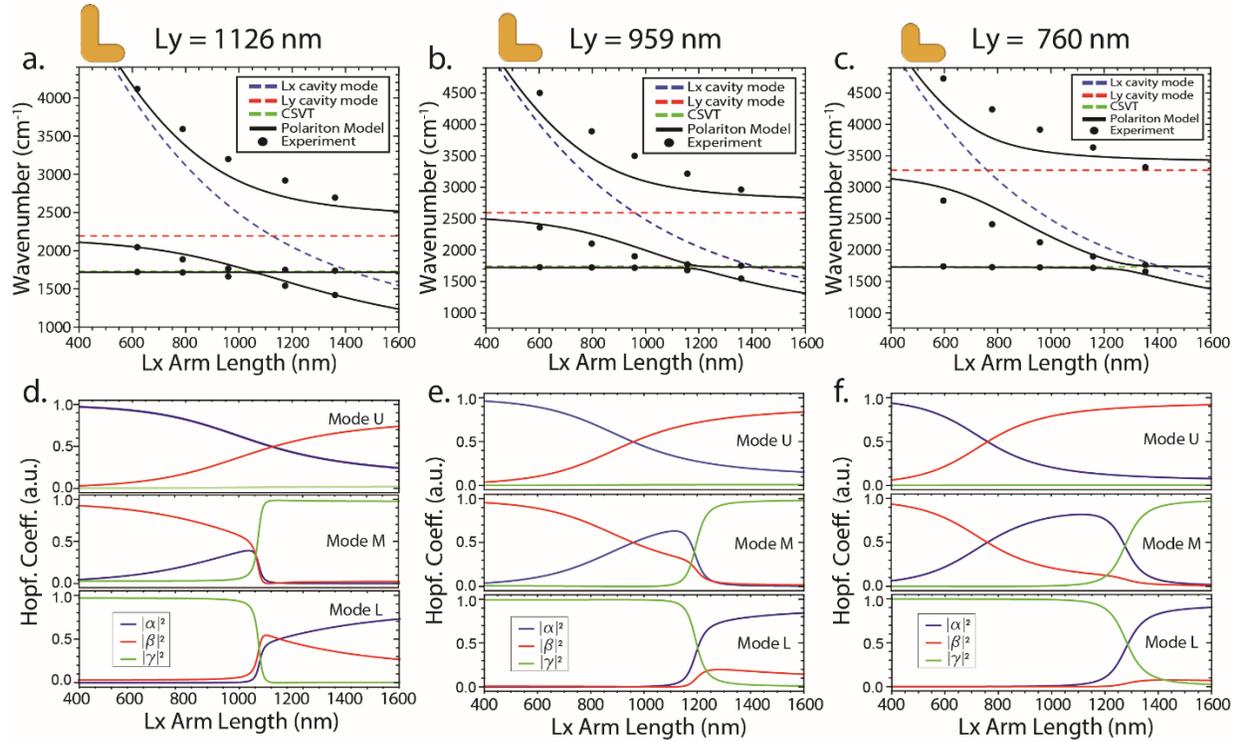

**Figure 4: Multimode coupling for various Ly arm lengths**. (a-c) The experimentally measured location of the upper, middle and lower polaritons (black circles) and the coupled three oscillator model (black lines) for three different average Ly arm lengths (a) 1126 nm, (b) 960 nm, (c) 760 nm. The uncoupled Lx (blue) & Ly (red) modes predicted by Eq. 1 and the vibrational transition (green) are given by the dashed lines. (d-f) The hopfield coefficents of the three coupled oscillator model in Eq. (3) as a function of Lx arm Length for three different average Ly arm lengths (d) 1126 nm, (e) 959 nm, (f) 760 nm.

To further understand the interplay between constituent modes ($\omega_x$, $\omega_y$, $\omega_{vib}$) and the behavior of the polariton branches, we apply our coupled oscillator model to several different Ly arm lengths using the $V_{1,2}$ values found in Figure 3 where *Ly* = 1300. Figure 4a shows the experimental and modeled mode dispersion as a function of *Lx* for several different *Ly* values, while SI Figure s8 shows the reflectance data. Focusing on the behavior of the L and M polaritons, as Ly decreases the anti-crossing point shifts to larger Lx arm length. The Rabi splitting between the L-M polaritons is smaller (15 cm$^{-1}$ for *Ly* = 1126 nm in Figure 4a) when the L-M anti-crossing point occurs for more symmetric L-cavities and then increases again as the L-M anti-crossing point occurs for more anisotropic L-cavity geometries (53 cm$^{-1}$ for *Ly* = 959 nm and 101 cm$^{-1}$ for *Ly* = 760 nm in figures 4b, c, respectively). The small Rabi splitting for *Ly* = 1126 occurs because $\omega_x$ and $\omega_y$ are resonant when $Ly = Lx = 1095$nm such that the U-M polaritons experience their anti-crossing for nearly the same conditions as the anti-crossing point for the L-M polaritons. As a result of $V_1 > V_2$, mode U applies 'pressure' to mode M and pushes it back towards mode L. Thus anisotropic L shapes produce larger Rabi splittings for the L-M polaritons in this particular regime where $\omega_y > \omega_{vib}$; this does not hold when $\omega_y < \omega_{vib}$. From how the resulting polariton modes, and in particular the L and M branches, depend on the three constituent plasmon and vibrational modes, we see that only through a multimode coupling model can the system be fully understood; the typical two-oscillator model would be insufficient.

From Figures 3c & 4(d-e), we can see how the character of each polariton branch changes as $Lx$ and $Ly$ are varied. For a decrease in the length $Ly$ (i.e. $\omega_y$ increases in energy) the U polariton crosses over in character from a primarily $\omega_x \to \omega_y$ mode, where the condition $\omega_y = \omega_x$ determines this crossover point. Furthermore, the contribution of $\omega_{vib}$ (although small) increases for longer $Ly$ lengths as $\omega_y$ generally provides the lower limit for the energy of U polariton in this regime and is thus allowed to approach closer to $\omega_{vib}$. For longer Ly arm lengths the M polariton goes from having mostly a $\omega_y \to \omega_{vib}$ nature as $Lx$ lengthens, and for shorter $Ly$ there is an increasing $\omega_x$ contribution such that the progression is mostly $\omega_y \to \omega_x \to \omega_{vib}$ in character. The change in $Ly$ has an almost converse effect on the character of the L polariton. Here, its character progresses through all three coupled modes ($\omega_{vib} \to \omega_y \to \omega_x$) for longer $Ly$ and then evolves from mostly $\omega_{vib} \to \omega_x$ at shorter $Ly$. The L and M polaritons show this behavior due to energy separation of $\omega_{vib}$ and $\omega_y$. When Ly is longer such that $\omega_y$ is closer in energy to $\omega_{vib}$, the Rabi splitting of the L-M polaritons results mostly from the interaction between these two modes. Here, $\omega_x$ plays a minor role, only contributing at longer $Lx$ when the energy difference between $\omega_x$ and the other resonances is small. For the case of a short $Ly$ such that $\omega_y$ is further in energy from $\omega_{vib}$, then the Rabi splitting of the L-M polaritons results mostly from the interaction between the two modes $\omega_x$ and $\omega_{vib}$, with $\omega_y$ contributing only to mode M for smaller $Lx$ where $\omega_x$ is detuned in energy from the other resonances. From the analysis above, we see that controlling the different aspect ratios of the L-cavities offers a way to tune the behavior and character of these modes.

## Discussion

Plasmon-vibrational coupling between the MIM L-cavity modes and the CSVT of the IP-Dip polymer fundamentally modified these two separate systems into a single new hybrid light-matter system with the three polariton eigenmodes given by Eq. 3. Evaluating the coupling strength of multiple coupled oscillators is less straightforward than the case of a simple two oscillator system. The Rabi splitting of 1101 cm$^{-1}$ between the U and L polariton bands clearly meets the proposed criteria[57] for strong multimode coupling as it is larger than the weighted average of the linewidths of the polariton branches, which is ~ 250 cm$^{-1}$. In fact, this meets most criteria for ultrastrong coupling because the Rabi splitting is on the order of the bare cavity frequency, the reduced coupling strength is >10%, and the geometric mean between the reduced coupling strength and cooperativity factor exceeds unity. However, the U-L Rabi splitting may not be representative of the light-matter coupling and could be better understood as dominated by plasmon-plasmon coupling as the U and L polaritons are dominated by plasmonic character for this arm length.

If we instead judge the plasmon-vibrational coupling based on the L-M Rabi splitting, because these modes retain the most vibrational mode character, then we see that the Rabi splitting is larger than the polariton line widths, so that the system may be better described by the strong coupling regime. A number of criteria for strong light-matter coupling are used in the literature based on the strength of the coupling interaction potential, the Rabi Splitting, and the line widths. The less strict criteria based on the simple two coupled oscillator model in the absence of damping is that twice the interaction potential be larger than the average line widths. However, merely meeting this criterion produces no visible Rabi splitting.

Since broadening reduces the observed Rabi splitting, the condition for visibility is that the Rabi splitting be larger than the average line widths.[26] The line width of modes L and M when they are primarily of CSVT and plasmonic character, respectively, are 33 cm$^{-1}$ and 163 cm$^{-1}$, respectively. Thus the averaged line width ( 98 cm$^{-1}$) is smaller than the Rabi splitting (104 cm$^{-1}$ and 101 cm$^{-1}$) for cavities where Ly=1300 nm and Ly=750 nm.[27] Therefore, these systems meet the criteria for strong light-matter coupling. So through this demonstration, we see that this 3D printed MIM L-shaped cavity can achieve strong plasmon-vibrational coupling if properly tuned with respect to a vibrational transition of interest.

The unique MIM L-cavity geometry employed here offers a number of advantages. The subwavelength confinement of the electromagnetic radiation and subsequent increase in the field intensity enables stronger coupling to the CSVT of the IP-Dip polymer with a reduced number of molecular vibrational oscillators as compared with an open cavity design.[43] Unlike e-beam lithography, the 3D printed nature of these cavities offers the ability to create complex cavity structures, such as creating nanoscopic voids in the insulator region for fluid flow similar to the microfluidic Fabry-Perot cavity scheme discussed in Ref [58]. So while the strong coupling to the CSVT of the Ip-Dip polymer used during fabrication may not itself have technological relevance, one can imagine schemes where strong coupling to other transitions is used for sensitive detection of analytes or to influence chemical reactions within solutions flowed through the 3D structure. The MIM L-cavities studied here have a volume of ~60 attoliters offering a platform for ultra-low volume sensing of unique vibrational fingerprints. The VSC causes the polariton dominated by vibrational character to borrow oscillator strength from the plasmon resonance, increasing its absorbance by 5-fold, a fact that could be exploited for chemical or biological detection. Strong coupling also enables the possibility to reversibly control the photochemical response and reaction rate of the material system inside the cavity[58] and unlike a reaction controlled by standard photochemistry, no photons are needed to modify a VSC-based chemical reaction as it occurs due to fluctuations in the vacuum field.[38] Additionally, as the MIM modes are angle independent and its dispersion is flat in momentum space, any chemical applications could occur at non-normal incidences, while other cavities with dispersion are limited to applications at normal incidences. [58,59] Nonlinear optical applications related to mid infrared transitions, such as intersubband transitions of low-dimensional semiconductors, may also be possible if such materials could be introduced into the cavity.[60]

The two plasmonic modes of interest in this study were the hybrid bonded and antibonded modes of the novel MIM L-cavity. We find that the polariton modes can be accurately modeled by the strongly coupled oscillator model in Eq. 3 where the contributing resonances of each arm are given by the relation in Eq. 1. While in this report, VSC required inclusion of a third oscillator associated with the CVST, in the absence of such light-matter coupling the L-cavity can be described using a two coupled-oscillator model that considers only cavity modes of each arm. The L structure is one of the simplest in-plane anisotropic structures and has been used to manipulate the polarization of light and chiral responses to circularly polarized light.[46–50] Because the plasmonic modes of the L-cavity are composed of the coupled orthogonal modes of the two arms, it is expected that the L-cavity could exhibit circular dichroic behavior when the arms are anisotropic.[61] This cavity geometry could therefore offer the potential to couple to chiral vibrational modes in order to selectively detect or modify the behavior of one stereoisomer within a chiral enantiomer pair. Therefore, identifying and understanding any chiral response from these or similar laser-printed structures is a future direction for this work.

## Conclusion

In summary, we demonstrated and characterized mid-IR multimode VSC in a nanoscale direct laser-printed MIM plasmonic nanogap cavity system. Such a fabrication scheme represents a cost-effective, highly accessible means of fabricating and rapidly evaluating mid-IR nanometer-sized plasmonic cavities which are capable of producing strong light-matter interactions in deeply subwavelength volumes. The utilized 3D printer technique could allow for complex cavity designs, including those with voids for introducing solution samples for applications based on microfluidic platforms. Upon analysis of the L-shaped cavity electromagnetic response via analytical and EM simulations, we find that the fundamental dipole modes are a superposition of responses from each arm and can be well-described by a two coupled oscillator model. The additional coupling to the CSVT of the polymer within the cavity requires a multimode coupled oscillator model, leading to the observed vibrational strong coupling between L and M polaritons and ultra-strong coupling between L/M and U polaritons. These multifaceted optical cavities potentially provide a way to modify and control a material's chemical reactivity via VSC in novel ways due to its anisotropic multi-modal optical response. For example, these structures could provide strong chiral fields to a particular chiral stereoisomer for the detection or selective chiral VSC-mediated chemistry. Additionally, this platform could be used for ultra-sensitive chemical detection across a wide bandwidth as the L-cavities support plasmon modes across the Mid-IR into the near-IR while maintaining a low mode volume.

## Acknowledgment

The authors thank Dr. Igor Vurgaftman for helpful discussions concerning strong coupling. This work was supported by the Office of Naval Research through core programs at the U.S. Naval Research Laboratory (NRL) and the NRL Nanoscience Institute. This research was performed while N.V.P., M.A.M., and N.S. held NRC Research Associateship awards at the U.S. Naval Research Laboratory.

## References

(1) Moskovits, M. Surface-Enhanced Spectroscopy. *Reveiws Mod. Phys.* **1985**, *57* (3), 783–823.

(2) Osawa, M. Surface-Enhanced Infrared Absorption. In *Near-Field Optics and Surface Plasmon Polaritons*; Springer Berlin Heidelberg: Berlin, Heidelberg, 2001; pp 163–187.

(3) Chen, C. K.; De Castro, A. R. B.; Shen, Y. R. Surface-Enchanced Second-Harmonic Generation. *Phys. Rev. Lett.* **1981**, *46* (2), 145–148.

(4) Gontijo, I.; Boroditsky, M.; Yablonovitch, E.; Keller, S.; Mishra, U. K.; DenBaars, S. P.; Krames, M. Coupling of InGaN Quantum Well Photoluminescence to Silver Surface Plasmons. *Phys. Rev. B* **1999**, *60* (16), 100–101.

(5) Kinsey, N.; Ferrera, M.; Shalaev, V. M.; Boltasseva, A. Examining Nanophotonics for Integrated Hybrid Systems: A Review of Plasmonic Interconnects and Modulators Using Traditional and Alternative Materials [Invited]. *J. Opt. Soc. Am. B* **2015**, *32* (1), 121.

(6) Tserkezis, C.; Esteban, R.; Sigle, D. O.; Mertens, J.; Herrmann, L. O.; Baumberg, J. J.; Aizpurua, J. Hybridization of Plasmonic Antenna and Cavity Modes: Extreme Optics of Nanoparticle-on-Mirror Nanogaps. *Phys. Rev. A - At. Mol. Opt. Phys.* **2015**, *92* (5), 1–6.

(7) Baumberg, J. J.; Aizpurua, J.; Mikkelsen, M. H.; Smith, D. R. Extreme Nanophotonics from Ultrathin Metallic Gaps. *Nature Materials*. 2019, pp 668–678.

(8) Fang, Y.; Sun, M. Nanoplasmonic Waveguides: Towards Applications in Integrated Nanophotonic Circuits. *Light: Science and Applications*. 2015, p 294.

(9) Nielsen, M. G.; Gramotnev, D. K.; Pors, A.; Albrektsen, O.; Bozhevolnyi, S. I. Continuous Layer Gap Plasmon Resonators. *Opt. Express* **2011**, *19* (20), 19310.

(10) Economou, E. N. Surface Plasmons in Thin Films. *Phys. Rev.* **1969**, *182* (2), 539.

(11) Bozhevolnyi, S. I.; Søndergaard, T. General Properties of Slow-Plasmon Resonant Nanostructures:

Nano-Antennas and Resonators. *Opt. Express* **2007**, *15* (17), 10869.

(12) Duffett, G.; Wirth, R.; Rayer, M.; Martins, E. R.; Krauss, T. F. Metal-Insulator-Metal Nanoresonators - Strongly Confined Modes for High Surface Sensitivity. *Nanophotonics* **2020**, *9* (6), 1547–1552.

(13) Liu, N.; Mesch, M.; Weiss, T.; Hentschel, M.; Giessen, H. Infrared Perfect Absorber and Its Application as Plasmonic Sensor. *Nano Lett.* **2010**, *10* (7), 2342–2348.

(14) Cattoni, A.; Ghenuche, P.; Haghiri-Gosnet, A. M.; Decanini, D.; Chen, J.; Pelouard, J. L.; Collin, S. Λ3/1000 Plasmonic Nanocavities for Biosensing Fabricated by Soft UV Nanoimprint Lithography. *Nano Lett.* **2011**, *11* (9), 3557–3563.

(15) Luo, Y.; Shepard, G. D.; Ardelean, J. V.; Rhodes, D. A.; Kim, B.; Barmak, K.; Hone, J. C.; Strauf, S. Deterministic Coupling of Site-Controlled Quantum Emitters in Monolayer WSe 2 to Plasmonic Nanocavities. *Nat. Nanotechnol.* **2018**, *13* (12), 1137–1142.

(16) Greffet, J. J.; Laroche, M.; Marquier, F. Impedance of a Nanoantenna and a Single Quantum Emitter. *Phys. Rev. Lett.* **2010**, *105* (11), 117701.

(17) Hill, M. T.; Marell, M.; P Leong, E. S.; Smalbrugge, B.; Zhu, Y.; Sun, M.; van Veldhoven, P. J.; Jan Geluk, E.; Karouta, F.; Oei, Y.-S.; Nötzel, R.; Ning, C.-Z.; Smit, M. K.; Zia, R.; Selker, M. D.; Catrysse, P. B.; Brongersma, M. L. Lasing in Metal-Insulator-Metal Sub-Wavelength Plasmonic Waveguides. *Opt. Express* **2009**, *17* (13), 11107.

(18) Rose, A.; Hoang, T. B.; McGuire, F.; Mock, J. J.; Ciracì, C.; Smith, D. R.; Mikkelsen, M. H. Control of Radiative Processes Using Tunable Plasmonic Nanopatch Antennas. *Nano Lett.* **2014**, *14* (8), 4797–4802.

(19) Huang, J.; Akselrod, G. M.; Ming, T.; Kong, J.; Mikkelsen, M. H. Tailored Emission Spectrum of 2D Semiconductors Using Plasmonic Nanocavities. *ACS Photonics* **2018**, *5* (2), 552–558.

(20) Sun, J.; Hu, H.; Pan, D.; Zhang, S.; Xu, H. Selectively Depopulating Valley-Polarized Excitons in Monolayer MoS2by Local Chirality in Single Plasmonic Nanocavity. *Nano Lett.* **2020**, *20* (7), 4953–4959.

(21) Weisbuch, C.; Nishioka, M.; Ishikawa, A.; Arakawa, Y. Observation of the Coupled Exciton-Photon Mode Splitting in a Semiconductor Quantum Microcavity. *Phys. Rev. Lett.* **1992**, *69* (23), 3314–3317.

(22) Agranovich, V. M.; Litinskaia, M.; Lidzey, D. G. Cavity Polaritons in Microcavities Containing Disordered Organic Semiconductors. *Phys. Rev. B* **2003**, *67* (8), 085311.

(23) Liu, X.; Galfsky, T.; Sun, Z.; Xia, F.; Lin, E. C.; Lee, Y. H.; Kéna-Cohen, S.; Menon, V. M. Strong Light-Matter Coupling in Two-Dimensional Atomic Crystals. *Nat. Photonics* **2014**, *9* (1), 30–34.

(24) Yu, X.; Yuan, Y.; Xu, J.; Yong, K. T.; Qu, J.; Song, J. Strong Coupling in Microcavity Structures: Principle, Design, and Practical Application. *Laser and Photonics Reviews*. 2018, pp 1–19.

(25) Kwek, L.-C.; Auffeves, A.; Gerace, D.; Richard, M.; Portolan, S.; Santos, M. F.; Miniature, C. *STRONG LIGHT- From Atoms to Solid-State Systems*; World Scientific Publishing Company, 2013.

(26) Khitrova, G.; Gibbs, H. M.; Kira, M.; Koch, S. W.; Scherer, A. Vacuum Rabi Splitting in Semiconductors. *Nature Physics*. 2006, pp 81–90.

(27) Törmö, P.; Barnes, W. L. Strong Coupling between Surface Plasmon Polaritons and Emitters: A Review. *Reports Prog. Phys.* **2015**, *78*, 013901.

(28) Baranov, D. G.; Wersäll, M.; Cuadra, J.; Antosiewicz, T. J.; Shegai, T. Novel Nanostructures and Materials for Strong Light-Matter Interactions. *ACS Photonics* **2018**, *5* (1), 24–42.

(29) Gómez, D. E.; Vernon, K. C.; Mulvaney, P.; Davis, T. J. Surface Plasmon Mediated Strong Exciton-Photon Coupling in Semiconductor Nanocrystals. *Nano Lett.* **2010**, *10* (1), 274–278.

(30) Long, J. P.; Simpkins, B. S. Coherent Coupling between a Molecular Vibration and Fabry-Perot Optical Cavity to Give Hybridized States in the Strong Coupling Limit. *ACS Photonics* **2015**, *2* (1), 130–136.

(31) Memmi, H.; Benson, O.; Sadofev, S.; Kalusniak, S. Strong Coupling between Surface Plasmon Polaritons and Molecular Vibrations. *Phys. Rev. Lett.* **2017**, *118* (12), 126802.

(32) Wang, J.; Yu, K.; Yang, Y.; Hartland, G. V.; Sader, J. E.; Wang, G. P. Strong Vibrational Coupling in Room Temperature Plasmonic Resonators. *Nat. Commun.* **2019**, *10* (1).

(33) Chervy, T.; Thomas, A.; Akiki, E.; Vergauwe, R. M. A.; Shalabney, A.; George, J.; Devaux, E.; Hutchison, J. A.; Genet, C.; Ebbesen, T. W. Vibro-Polaritonic IR Emission in the Strong Coupling Regime. *ACS Photonics* **2018**, *5* (1), 217–224.

(34) Dunkelberger, A. D.; Spann, B. T.; Fears, K. P.; Simpkins, B. S.; Owrutsky, J. C. Modified Relaxation Dynamics and Coherent Energy Exchange in Coupled Vibration-Cavity Polaritons. *Nat. Commun.* **2016**, *7*.

(35) Ahn, W.; Vurgaftman, I.; Dunkelberger, A. D.; Owrutsky, J. C.; Simpkins, B. S. Vibrational Strong Coupling Controlled by Spatial Distribution of Molecules within the Optical Cavity. *ACS Photonics* **2018**, *5* (1), 158–166.

(36) Folland, T. G.; Lu, G.; Bruncz, A.; Nolen, J. R.; Tadjer, M.; Caldwell, J. D. Vibrational Coupling to Epsilon-Near-Zero Waveguide Modes. *ACS Photonics* **2020**, *7* (3), 614–621.

(37) Joseph, K.; Kushida, S.; Smarsly, E.; Ihiawakrim, D.; Thomas, A.; Paravicini-Bagliani, G. L.; Nagarajan, K.; Vergauwe, R.; Devaux, E.; Ersen, O.; Bunz, U. H. F.; Ebbesen, T. W. Supramolecular Assembly of Conjugated Polymers under Vibrational Strong Coupling. *Angew. Chemie - Int. Ed.* **2021**, *60* (36), 19665–19670.

(38) Hutchison, J. A.; Schwartz, T.; Genet, C.; Devaux, E.; Ebbesen, T. W. Modifying Chemical Landscapes by Coupling to Vacuum Fields. *Angew. Chemie - Int. Ed.* **2012**, *51* (7), 1592–1596.

(39) Thomas, A.; George, J.; Shalabney, A.; Dryzhakov, M.; Varma, S. J.; Moran, J.; Chervy, T.; Zhong, X.; Devaux, E.; Genet, C.; Hutchison, J. A.; Ebbesen, T. W. Ground-State Chemical Reactivity under Vibrational Coupling to the Vacuum Electromagnetic Field. *Angew. Chemie - Int. Ed.* **2016**, *55* (38), 11462–11466.

(40) Braun, A.; Maier, S. A. Versatile Direct Laser Writing Lithography Technique for Surface Enhanced Infrared Spectroscopy Sensors. *ACS Sensors* **2016**, *1* (9), 1155–1162.

(41) Verdelli, F.; Schulpen, J. J. P. M.; Baldi, A.; Rivas, J. G. Chasing Vibro-Polariton Fingerprints in Infrared and Raman Spectra Using Surface Lattice Resonances on Extended Metasurfaces. *J. Phys. Chem. C* **2022**, *19*, 18.

(42) Menghrajani, K. S.; Nash, G. R.; Barnes, W. L. Vibrational Strong Coupling with Surface Plasmons and the Presence of Surface Plasmon Stop Bands. *ACS Photonics* **2019**, *6* (8), 2110–2116.

(43) Dayal, G.; Morichika, I.; Ashihara, S. Vibrational Strong Coupling in Subwavelength Nanogap Patch Antenna at the Single Resonator Level. *J. Phys. Chem. Lett.* **2021**, *12* (12), 3171–3175.

(44) Deubel, M.; Von Freymann, G.; Wegener, M.; Pereira, S.; Busch, K.; Soukoulis, C. M. Direct Laser Writing of Three-Dimensional Photonic-Crystal Templates for Telecommunications. *Nat. Mater.* **2004**, *3* (7), 444–447.

(45) Dietrich, P. I.; Blaicher, M.; Reuter, I.; Billah, M.; Hoose, T.; Hofmann, A.; Caer, C.; Dangel, R.;

Offrein, B.; Troppenz, U.; Moehrle, M.; Freude, W.; Koos, C. In Situ 3D Nanoprinting of Free-Form Coupling Elements for Hybrid Photonic Integration. *Nat. Photonics* **2018**, *12* (4), 241–247.

(46) Sung, J.; Hicks, E. M.; Van Duyne, R. P.; Spears, K. Nanoparticle Spectroscopy: Plasmon Coupling in Finite-Sized Two-Dimensional Arrays of Cylindrical Silver Nanoparticles. *J. Phys. Chem. C* **2008**, *112* (11), 4091–4096.

(47) Lévesque, Q.; Makhsiyan, M.; Bouchon, P.; Pardo, F.; Jaeck, J.; Bardou, N.; Dupuis, C.; Haïdar, R.; Pelouard, J. L. Plasmonic Planar Antenna for Wideband and Efficient Linear Polarization Conversion. *Appl. Phys. Lett.* **2014**, *104* (11), 111105.

(48) Yang, J.; Zhang, J. Nano-Polarization-Converter Based on Magnetic Plasmon Resonance Excitation in an L-Shaped Slot Antenna. *Opt. Express* **2013**, *21* (7), 7934.

(49) Blanchard, R.; Aoust, G.; Genevet, P.; Yu, N.; Kats, M. A.; Gaburro, Z.; Capasso, F. Modeling Nanoscale V-Shaped Antennas for the Design of Optical Phased Arrays. *Phys. Rev. B* **2012**, *85* (15), 155457.

(50) Black, L. J.; Wang, Y.; De Groot, C. H.; Arbouet, A.; Muskens, O. L. Optimal Polarization Conversion in Coupled Dimer Plasmonic Nanoantennas for Metasurfaces. *ACS Nano* **2014**, *8* (6), 6390–6399.

(51) Fullager, D. B.; Boreman, G. D.; Hofmann, T. Infrared Dielectric Response of Nanoscribe IP-Dip and IP-L Monomers after Polymerization from 250 Cm−1 to 6000 Cm−1. *Opt. Mater. Express* **2017**, *7* (3), 888.

(52) Li, H. H. Refractive Index of Silicon and Germanium and Its Wavelength and Temperature Derivatives. *J. Phys. Chem. Ref. Data* **1980**, *9* (3), 561–658.

(53) Palik, E. D. *Handbook of Optical Constants of Solids*; 1997.

(54) Black, L. J.; Wang, Y.; De Groot, C. H.; Arbouet, A.; Muskens, O. L. Optimal Polarization Conversion in Coupled Dimer Plasmonic Nanoantennas for Metasurfaces. *ACS Nano* **2014**, *8* (6), 6390–6399.

(55) Savona, V.; Andreani, L. C.; Schwendimann, P.; Quattropani, A. Quantum Well Excitons in Semiconductor Microcavities: Unified Treatment of Weak and Strong Coupling Regimes. *Solid State Commun.* **1995**, *93* (9), 733–739.

(56) Deshmukh, R.; Marques, P.; Panda, A.; Sfeir, M. Y.; Forrest, S. R.; Menon, V. M. Modifying the Spectral Weights of Vibronic Transitions via Strong Coupling to Surface Plasmons. *ACS Photonics* **2020**, *7* (1), 43–48.

(57) Liu, Y.; Zhu, Z.; Qian, J.; Yuan, J.; Yan, J.; Shen, Z. X.; Jiang, L. Strong Coupling between Two-Dimensional Transition Metal Dichalcogenides and Plasmonic-Optical Hybrid Resonators. *Phys. Rev. B* **2021**, *104* (20), 1–8.

(58) Nagarajan, K.; Thomas, A.; Ebbesen, T. W. Chemistry under Vibrational Strong Coupling. *Journal of the American Chemical Society*. 2021, pp 16877–16889.

(59) Li, T. E.; Nitzan, A.; Subotnik, J. E. Collective Vibrational Strong Coupling Effects on Molecular Vibrational Relaxation and Energy Transfer: Numerical Insights via Cavity Molecular Dynamics Simulations. *Angew. Chemie - Int. Ed.* **2021**, *60* (28), 15533–15540.

(60) Lee, J.; Tymchenko, M.; Argyropoulos, C.; Chen, P. Y.; Lu, F.; Demmerle, F.; Boehm, G.; Amann, M. C.; Alù, A.; Belkin, M. A. Giant Nonlinear Response from Plasmonic Metasurfaces Coupled to Intersubband Transitions. *Nature* **2014**, *511* (7507), 65–69.

(61) Kong, X. T.; Khosravi Khorashad, L.; Wang, Z.; Govorov, A. O. Photothermal Circular Dichroism Induced by Plasmon Resonances in Chiral Metamaterial Absorbers and Bolometers. *Nano Lett.* **2018**, *18* (3), 2001–2008.

Supplementary Information

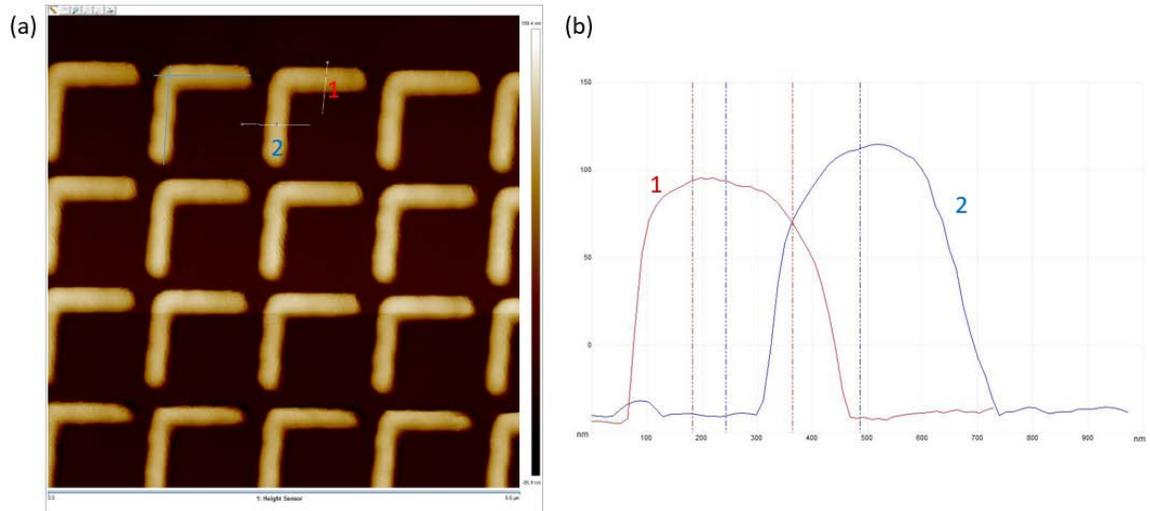

**Supplementary Figure 1 | AFM characterization of select L antennas** (a) AFM image of select L Antennas with 70nm of evaporated Au on top. The arm lengths are 675 nm in both directions and is from the same sample as the SEM images in Figure 1c. (b) Cut-lines labeled in (a) across each arm.

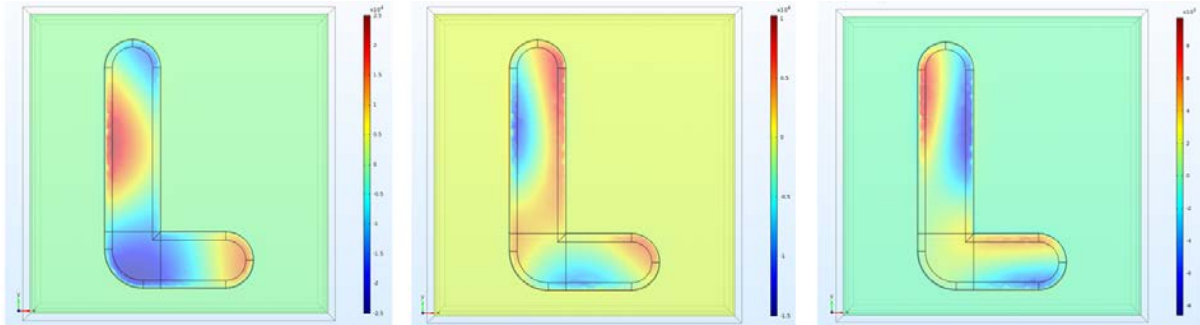

**Supplementary Figure 2 | Simulated reflectance spectra for L-antenna in Figure 2** –The field mode plots of higher-order modes are not characterized (5750)/ not seen (4400,6050) in the experimental data.

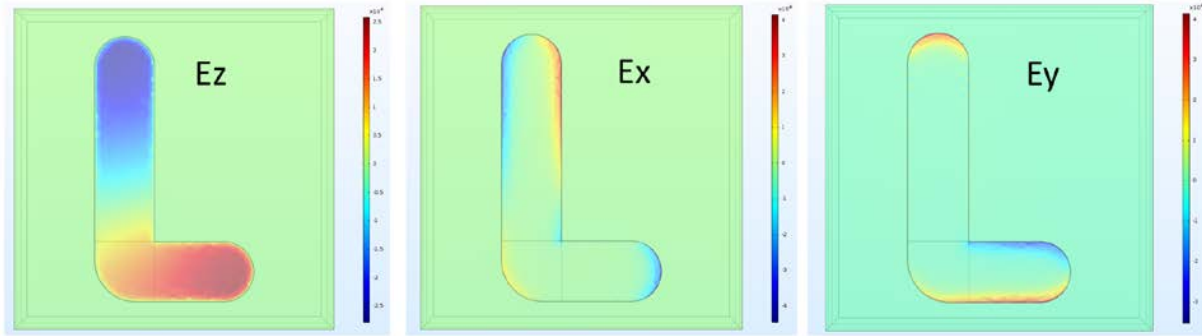

**Supplementary Figure 3 | E-field components of Mode L in Figure 2** –Electric field mode (Ez, Ex, Ey) plots of Mode A Lower, Ez also shown in Figure 2a.

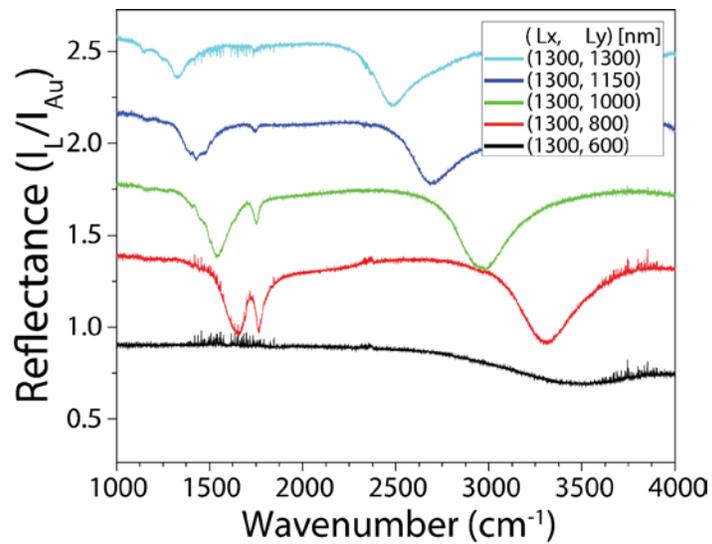

**Supplementary Figure 4 | Size-dependent dispersion of the L-antenna modes -** Reflectance spectra of several different sized L-antennas. Here Ly is varied increments of ~200 nm while Lx is kept nearly constant at ~1300 nm. The modal dispersion is similar to Figure 4a where Lx was varied instead of Ly. The Ly distances given in the legend are nominal values.

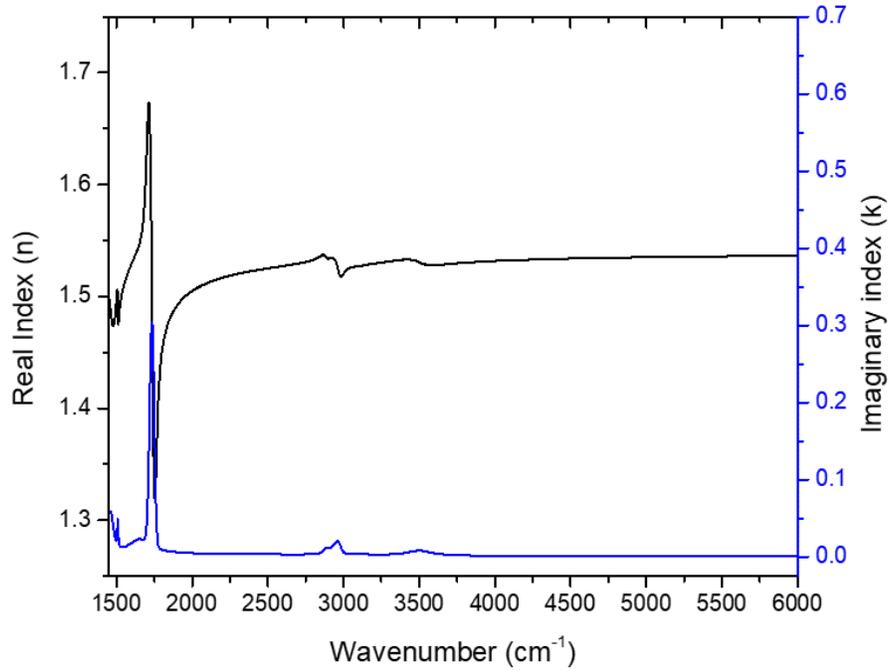

**Supplementary Figure 5 | Ellipsometry data of Ip-dip**. The experimentally measured real (black) and Imaginary (blue) parts of the index of refraction for Ip-dip polymer (black) taken from Ref 48 of the main text.

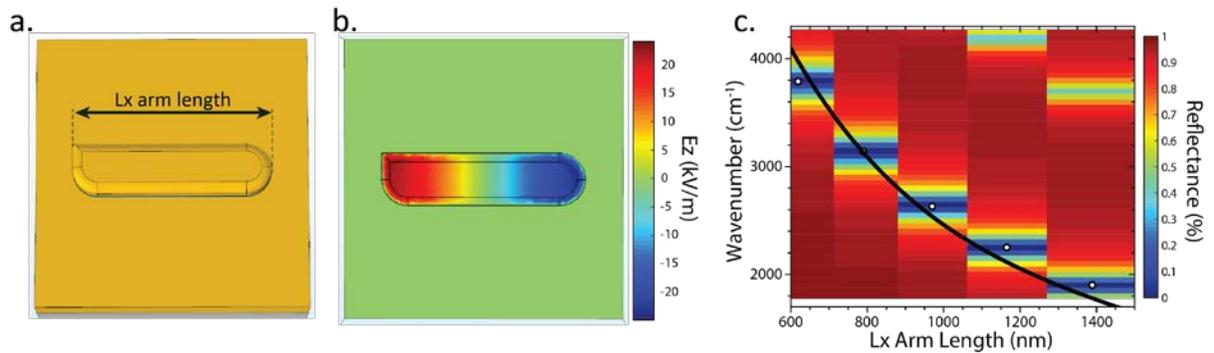

**Supplementary Figure 6 | Size-dependent modal dispersion of a single arm** (a) Geometry of a single arm (Lx=1165 nm) used to simulate the dispersion in the contour plot of (b). (b) The field mode profile at 2250 cm$^{-1}$ for a Lx arm length=1165 nm. (c) Simulated reflectance spectra of the structures in (a) & (b) shown as a contour plot overlayed with the calculated dispersion of Eq. 1 of the main text.

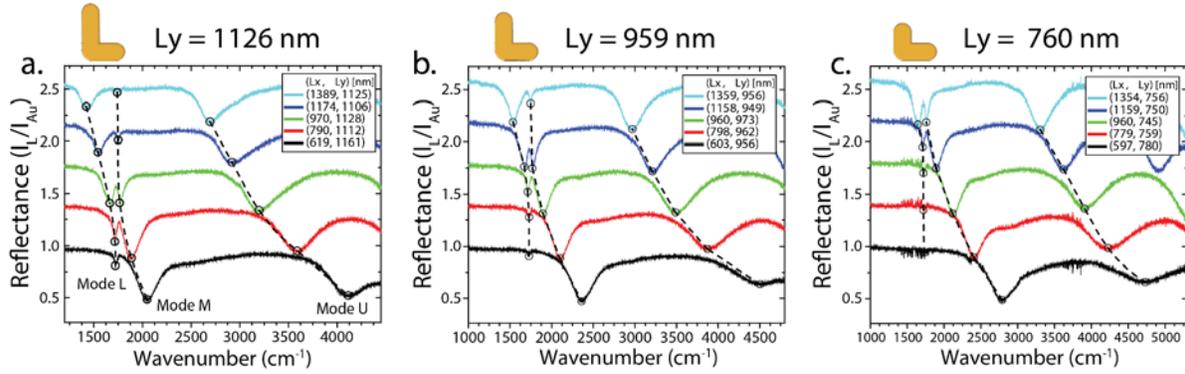

**Supplementary Figure 8 | Vibrational-Plasmon coupling for various Ly**. (a-c) Waterfall plots of the experimental reflectance as Lx is varied for constant L of modes L, M & U for three different Ly arm lengths (a) 1126 nm, (b) 959 nm, (c) 760 nm.

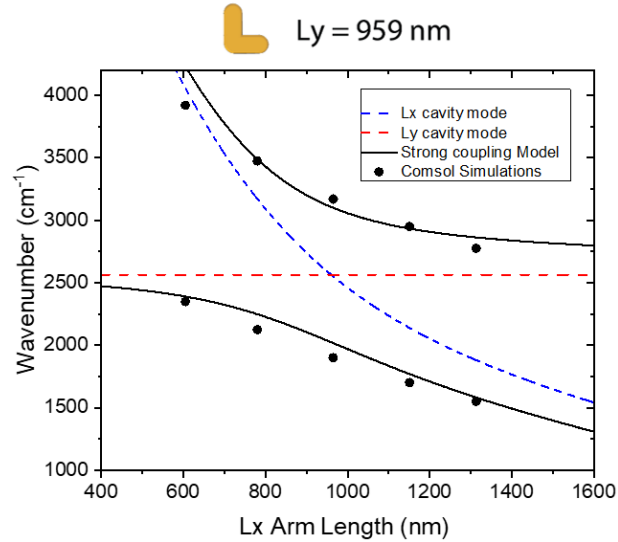

**Supplementary Figure 7 | Modeled L-cavity bonded and anti-bonded modes without the presence of the Ip-dip vibrational transitions.** The L-cavity modes were simulated via COMSOL for various Lx arm lengths while the Ly arm is kept constant at 959 nm. The black solid lines are the bonded (lower) and anti-bonded (upper) modes modeled via a 2-oscillator strong coupling model of the two uncoupled Lx and Ly cavity resonances associated with each arm (dashed lines). For this, Eq. (3) is modified by setting Plasmon-CSVT interaction potential to 0, $V_2=0$ and removing the CSVT, $\omega_{vib}$. Thus the eigenvalue problem of eq. 3 is given now by 2x2 matrix, $\begin{bmatrix} \omega_x[x] & V_1 \\ V_1 & \omega_y \end{bmatrix} \begin{bmatrix} \alpha[x] \\ \beta[x] \end{bmatrix} = \lambda_{\pm} \begin{bmatrix} \alpha[x] \\ \beta[x] \end{bmatrix}$, where $\lambda_{\pm}$ are now the two eigenvalues of the system. The plasmon-plasmon interaction potential, $V_1$, was kept at 541 cm$^{-1}$ as was found in figures 3 & 4 of the main text.